\documentclass[fleqn,twoside]{article}
\usepackage{espcrc2}

\usepackage{graphicx}

\usepackage{epsfig}
\usepackage{amsfonts}
\newcommand{\AmS}{{\protect\the\textfont2
  A\kern-.1667em\lower.5ex\hbox{M}\kern-.125emS}}

\newcommand{\vslash}{\ensuremath{v\!\!\! /}}

\def\half{{\scriptstyle \raise.2ex\hbox{${1\over2}$}}}
\def\fourth{{\scriptstyle \raise.2ex\hbox{${1\over4}$}}}

\newcommand*{\chpt}{\raise0.4ex\hbox{$\chi$}PT}
\newcommand*{\schpt}{S\raise0.4ex\hbox{$\chi$}PT}
\newcommand*{\ie}{\textit{i.e.},\ }

\newcommand*{\et}{\textit{et al.}}

\newcommand*{\GeV}{{\rm Ge\!V}}

\newcommand{\trD}{\ensuremath{\textrm{tr}_{\textrm{\tiny \it D}}}}

\newcommand{\cL}{\mathcal{L}}
\newcommand{\cM}{\mathcal{M}}

\newcommand{\cO}{\mathcal{O}}
\newcommand{\cP}{\mathcal{P}}

\newcommand{\cV}{\mathcal{V}}

\def\eq#1{Eq.~(\ref{eq:#1})}

\title{Staggered Chiral Perturbation Theory with Heavy-Light
Mesons}

\author{C. Aubin, C. Bernard
\address{Department of Physics, 
Washington University, St. Louis, MO 63130, USA}}

\begin{document}

\begin{abstract}
  We merge heavy quark effective theory with staggered chiral perturbation
  theory to calculate heavy-light ($B$, $D$) meson quantities. 
  We present results at NLO for the $B(D)$ meson decay
  constant in the partially quenched and full QCD cases, and
  discuss the calculation of the form factors for $B(D)\to \pi(K) \ell \nu$
  decays.
\end{abstract}

\maketitle

The lattice can make a major contribution to the understanding of
flavor physics through the computation of the properties of
heavy-light mesons (see Ref.~\cite{Kronfeld:2003sd} for 
reviews). A promising approach for these systems is to use staggered light
quarks so that the simulations can be performed in the chiral regime 
\cite{Wingate:2002fh,Aubin:2004ej}.
It is therefore important to understand the effects of staggered taste violations on heavy-light
quantities. To do so, we contruct the
effective theory of heavy-light mesons and the light pseudoscalar mesons
(collectively referred to as ``pions'' here) that is a merging of
Heavy Quark Effective Theory (HQET) \cite{Burdman:1992gh}
and Staggered Chiral Perturbation
Theory (\schpt) \cite{Lee:1999zx,Aubin:2003mg}.

For the lattice simulations we wish to describe, the heavy quark
mass $m_Q$ is large but not infinite.  The fact that $m_Q$ is large 
($\gg \Lambda_{\rm QCD}$) means that neglecting $1/m_Q$ terms 
in our effective theory is a reasonable first approximation.  
The fact that  $m_Q$ is not infinite
is, surprisingly, also important in simplifying the theory.
The criterion is 
$\Delta \! E(\pi/a)\equiv [(m_Q^2 + (\pi/a)^2]^{\frac{1}{2}}-m_Q \gg 
m_\pi$. 
This guarantees that if a light quark changes its taste by exchanging
a gluon of momentum $\pi/a$ with a heavy quark, the latter gets
so much energy that it decouples from the chiral theory.
A worst case example is a $b$ quark ($m_Q\sim 5\,\GeV$) on a ``coarse''
lattice ($a \approx 0.125\,$fm;  $\pi /a \approx 5\GeV$). Here,
$\Delta \! E(\pi/a)\approx 2\,\GeV$, which is safely higher than physical
or simulated pion masses. Heavy quarks with energy this high
affect low-energy physics only though renormalization.

Technically, the criterion $\Delta \! E(\pi/a)\gg m_\pi$ allows us
to neglect, from the effective Symanzik action of the lattice theory,
``mixed'' 4-quark operators (products of heavy and light bilinears)
that violate taste.  At $\cO(a^2)$, then, all taste-violations in
the Symanzik action are in the light quark sector, \ie the same
as in Refs.~\cite{Lee:1999zx,Aubin:2003mg}.

We stress that we are not taking into account discretization
effects due to the heavy quarks, 
only those that arise specifically
from light-quark taste violations. Ideally,
with a highly improved heavy quark, 
the former effects would be negligible. With currently used 
heavy quark actions,
these effects are not negligible and must be estimated
and/or extrapolated away separately.

Once the Symanzik action is known,
combining HQET and \schpt\ is a straightforward generalization
of continuum heavy-light \chpt\ \cite{MANOHAR-WISE}. 
The heavy-light mesons are combined into a single field and its conjugate
\begin{equation}
  H_a  = \frac{1 + \vslash}{2}\left[ \gamma^\mu B^{*}_{\mu a}
    - \gamma_5 B_{a}\right]\ , \
  \overline{H}_a   \equiv   \gamma_0 H^{\dagger}_a\gamma_0
  \end{equation}
while the light mesons are collected in $\Sigma = \sigma^2 = \exp
(i\Phi/f)$. $\Phi$ is the $12\times 12$ matrix 
(for 3 flavors of light quarks)
that contains the pions
 \begin{eqnarray}
      \Phi = \left( \begin{array}{ccc}
     	U  & \pi^+ & K^+ \\
     	\pi^- & D & K^0  \\
	K^-  & \bar{K^0}  & S  \end{array} \right),
  \end{eqnarray}
where $U = U_a T_a$, $K^+ = K^+_a T_a$, etc., and the $T_a$ are the
16 taste matrices \cite{Lee:1999zx,Aubin:2003mg}. 
Under chiral $SU(12)_L\times SU(12)_R$ we
have the transformations
\begin{eqnarray}
  H_a \to  H_b U^{\dagger}_{ba}&\qquad &
  \overline{H}_a \to  U_{ab}\overline{H}_b \\
  \sigma \to  L\sigma U^{\dagger} = U \sigma R^{\dagger}&\qquad &
  \Sigma \to  L\Sigma R^{\dagger} , 
\end{eqnarray}
with $L,R,U\in SU(12)$.

To leading order in $1/m_Q$, there are three expansion parameters:
$m_\pi\sim \sqrt{m_q}$ ($m_q$ is a generic light quark mass), $a^2$,
and $k$, the heavy-light residual momentum. We assume $k\sim m_\pi$
and $m_q\sim a^2$ for our power counting.
We write the Lagrangian as $\cL = \cL_{1,2} + \cL_{3}$, where 
\begin{eqnarray}
  \cL_{1,2} & = &  i \trD[\overline{H}_a v^{\mu}(\delta_{ab}
    \partial_{\mu}+ i \mathbb{V}^{ba}_{\mu}) H_b]
  \nonumber\\&&{}
  + g_\pi \trD(\overline{H}_a H_b\gamma^{\nu}\gamma_5 \mathbb{A}^{ba}_{\nu})
    + \cL_{\schpt}\ .
%
\end{eqnarray}
$\cL_{\schpt}$ is the pion \schpt\ Lagrangian found in
Ref.~\cite{Aubin:2003mg}. 
$\cM$ is the light quark mass matrix, 
  $\mathbb{V}_{\mu}  \equiv  (i/2) \left[ \sigma^{\dagger} \partial_\mu
    \sigma + \sigma \partial_\mu \sigma^{\dagger}   \right]$, and
$ \mathbb{A}_{\mu}  \equiv  (i/2) \left[ \sigma^{\dagger} \partial_\mu
    \sigma - \sigma \partial_\mu \sigma^{\dagger}   \right]$.

The terms in $\cL_{1,2}$ contribute to the NLO chiral logarithms,
while the terms in $\cL_3$ are one order higher in the
calculation, and thus can at most contribute to analytic terms at NLO. 
Part of $\cL_3$ is a new taste-breaking potential, $\cV_H$, containing
both heavy and light mesons \cite{AubinBernard}. 

Decay constants can be extracted from the matrix element
$  \left\langle 0 \left| A^\mu_{x,a}
  \right| B_{x,b}(v) \right\rangle =-i
  f_{B_{x}}m_{B_{x}} v^{\mu} \delta_{ab}$,
where $A^\mu_{x,a}$ 
is the axial current which destroys a $B_{x,a}$ meson, $x$
denotes the light quark flavor, and $a$, its taste. The decay constant
$f_{B_x}$ is independent of the light quark taste due to
the symmetry $\sigma\to
\xi^{(3)}_\mu\sigma\xi^{(3)}_\mu$, $H \to H\xi^{(3)}_\mu $.
The corresponding chiral operator is 
$A^\mu_{x,a} = \frac{i\kappa}{2}
\trD\left[\gamma^\mu (1-\gamma_5)(\cP_x H_{b})\right]\sigma_{ba}
$, where $\cP_x$ projects out the $x$ flavor block of $H_{b}$.
We write the decay constant as
$f_{B_{x}}\sqrt{m_{B_x}} = \kappa\left( 1 + \delta\! f_{B_{x}} 
/(16\pi^2 f^2) \right)$.

\begin{figure}[t]
  \centering
  \includegraphics[width=2.8in]{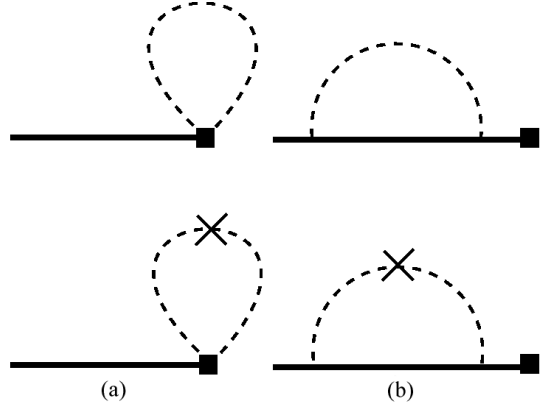}
\vspace{-0.9cm}
  \caption{One-loop diagrams contributing to the (a) current
  correction and (b) wavefunction renormalization.}
  \label{fig:tadF}
\vspace{-0.3cm}
\end{figure}

There are two types of non-zero one-loop diagrams which contribute to the 
chiral logarithms, one coming from corrections to the
current itself [Fig.~\ref{fig:tadF}(a)] and the other from 
wavefunction renormalization [Fig.~\ref{fig:tadF}(b)]. The
crosses in Fig.~\ref{fig:tadF} 
refer to one or more insertion of the two-point hairpin
diagrams discussed in Ref.~\cite{Aubin:2003mg}, 
for the singlet, axial, and vector taste
flavor-neutral mesons. These correspond to disconnected quark level
diagrams. 
We account for the transition from four to one tastes per flavor in
the same way as in the calculations for light meson
quantities \cite{Aubin:2003mg}. 

For the $1\!+\!1\!+\!1$ partially quenched chiral logs, we 
define the following sets of masses,
\begin{eqnarray}
  \mu_t^{(3)} & = & \{m_{U_t}^2,m_{D_t}^2 ,m_{S_t}^2 \}\ , \nonumber \\
 M_I^{(3)} & = & \{m_{X_I}^2, m_{\pi^0_I}^2,m_{\eta_I}^2  \}\ ,\nonumber \\
   M_V^{(4)} & = &  \{m_{X_V}^2, m_{\pi^0_V}^2,m_{\eta_V}^2,m_{\eta'_V}^2  \}\ ,
\end{eqnarray}
where $t$ is a taste label.
We then obtain
\begin{eqnarray}\label{eq:1p1p1_pq_fB}
  \delta\! f_{B_x}
  \!\!&\!\!\!\! =\!\!\!\! & \!\!  -\frac{1+3g_\pi^2}{2}
  \biggl\{\frac{1}{16}\sum_{F,t} \ell(m_{(xF)_t}^2)
  \nonumber\\*\!\!&\!\!\!\!\!\!\!\!&\!\!{}+
  \frac{1}{3}\sum_{j_I}
  \partial_{X_I}
       \left[ R^{[3,3]}_{j_I}( M_I^{(3)};
	  \mu_I^{(3)}) \ell(m_{j_I}^2)\right]
  \nonumber\\*\!\!&\!\!\!\!\!\!\!\!&\!\!{} + a^2\delta'_V 
    \!\!\sum_{j_V}
    \partial_{X_V}\!\!\!
    	\left[
    R^{[4,3]}_{j_V}
    ( M_V^{(4)};
	 \mu_V^{(3)})\ell(m_{j_V}^2)\right]\nonumber
	 \\*\!\!&\!\!\!\!\!\!\!\!&\!\!{}
  + [V\to A] 
  \biggr\}    \, ,
\end{eqnarray}
where $\ell(m^2) = m^2 \ln m^2 +$finite volume corrections, $F$ labels
sea-quark flavors, $t$ runs over the 16 tastes,
and $j_I$ and $j_V$ run over the set of masses in the first argument 
of the $R_j$, the residues of the poles of the disconnected
flavor-neutral propagators 
\cite{Aubin:2003mg}. We define the derivative in
\eq{1p1p1_pq_fB}  by
$\partial_X \equiv \partial/\partial m^2_X$. 

In 
the $2\!+\!1$ ($m_u = m_d \ne m_s$) full QCD case,
we have 
\begin{eqnarray}
  \delta\! f_{B}
 \! \!\!\!\!\!& = &\!\!\!\!\! -\frac{1+3g_\pi^2}{2}
  \biggl\{\frac{1}{16}\sum_{t} \left[ 2\ell(m_{\pi_t}^2) 
    + \ell(m_{K_t}^2) \right]    \nonumber \\*&&{}\!\!\!\!\!-
  \frac{1}{2}\ell(m^2_{\pi^0_I}) + \frac{1}{6}\ell(m^2_{\eta_I})  
    \nonumber \\*&&{}
    \!\!\!\! \!\!\!\!\!+   a^2\delta'_V \biggl[ 
     \frac{(m^2_{\pi^0_V} - m^2_{S_V})\ell(m_{\pi^0_V}^2)}{(m^2_{\pi^0_V} - 
       m^2_{\eta_V})(m^2_{\pi^0_V} - m^2_{\eta'_V})}
          \nonumber \\*&&{}\!\!\!\!\!\!\!\!\!+ (\pi^0_V\to\eta_V\to\eta'_V\to\pi^0_V)
     \biggr] + [V\to A] 
   \biggr\},
\end{eqnarray}
where $(\pi^0_V\to\eta_V\to\eta'_V\to\pi^0_V)$ represents two additional
terms with cyclic replacements.
There is a similar result for $\delta\!f_{B_s}$.

\begin{figure}
  \centering
  \includegraphics[width=2.8in]{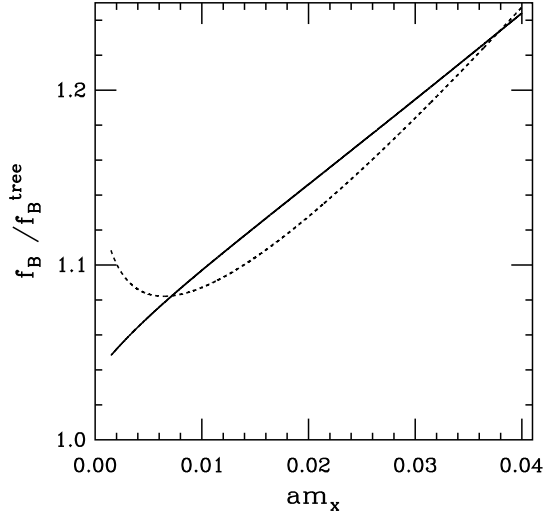}
\vspace{-0.9cm}
  \caption{Ratio of the one-loop to the tree-level
    partially quenched $B$ decay constant for 
    2 degenerate sea quarks ($am_{\rm sea} = 0.010$) with (solid line) 
    and without (dashed line)
    the lattice terms.}
  \label{fig:fB_am010_both}
\vspace{-0.5cm}
\end{figure}

To see the importance of the additional $a^2$ terms, 
we compare, 
in Fig.~\ref{fig:fB_am010_both}, the continuum version
of the chiral logs as a function of valence quark mass with 
the complete lattice expression. 
The $a^2$ 
terms drastically change the behavior for lighter valence
quark masses. We conclude that lattice data for
heavy-light decay constants can be misleading; the data may look linear,
but there can be large systematic errors if
physical values are extracted by simple linear extrapolation.

Comparing our results for decay constants to the continuum
expressions, we see that 
there is an easy way to generalize the continuum, partially
quenched \chpt\ expressions with $N_{\rm sea}$ degenerate quark
flavors to \schpt.
Terms $\propto N_{\rm sea}$, which arise from connected diagrams that
at the quark level involve sea quark loops, become an
average over tastes. Terms $\propto 1/N_{\rm sea}$ are
disconnected; in \schpt\ we have $I, V,$ and $A$ pieces.
There can be additional minus signs in disconnected
$A$ and $V$ terms compared to disconnected $I$ terms,
because of the anti-commutation relations for taste matrices. 
With this in mind, it is straightforward to write down the
\schpt\ expressions for the form factors for $B(D)\to\pi (K)\ell\nu$
decays using Ref.~\cite{Becirevic:2003ad}. 
These expressions are however quite complicated and will be presented
separately \cite{AubinBernard}.
It will also be straightforward to 
extend these results to heavy-light B-parameters. 

Fortunately, at this order the heavy-light \schpt\ chiral logarithms
require no new parameters beyond what are already present in
continuum heavy-light \chpt\ and light meson \schpt.  There will
however be simple new analytic terms, proportional to $a^2$, in
the complete expressions.  
Our results have already been
used in analyzing lattice data for heavy-light form-factors and decay
constants \cite{Aubin:2004ej}.  

This work was supported by the U.S.\ DOE.

\end{document}